\documentclass{aastex631}
\usepackage{bm}
\usepackage{subcaption}
\usepackage[T1]{fontenc}
\usepackage{booktabs}
\usepackage{threeparttable}
\usepackage{multirow}
\usepackage{amsmath}
\usepackage[utf8]{inputenc}
\usepackage{makecell}
\usepackage{numprint}
\usepackage[ruled,vlined]{algorithm2e}
\usepackage{tabularx}

\begin{document}

 \title{A Galton Board Approximation Method for Estimating Pedestrian Walking Preferences within Crowds}
 
\author{Jinghui Wang}

\affiliation{School of Safety Science and Emergency Management\\
Wuhan University of Technology\\
Wuhan, China}

\author{Wei Lv}
\altaffiliation{{\url{weil@whut.edu.cn}}}
\affiliation{School of Safety Science and Emergency Management\\
Wuhan University of Technology\\
Wuhan, China}

\begin{abstract}

This paper proposes a Galton board approximation method to analyze the potential walking preferences of pedestrians. We employ the binomial distribution to estimate the walking preferences of pedestrians in dynamic crowds. Estimating the probability of the right-side preference (\(p\)) based on observed data poses the challenge, as statistical measures such as means and variances often lead to divergent results. This study aims to explore this issue.

\end{abstract}

\section{Data} 
\label{section1}
Pedestrian walking preferences are influenced by individual habits and social norms, making quantitative statistical analysis in dense crowds highly challenging \citep{moussaid2009experimental,ma2010experimental}. We have conceived a Galton board approximation method to quantitatively estimate pedestrian walking preferences.

\begin{figure}[ht!]
\centering
\includegraphics[scale=0.35]{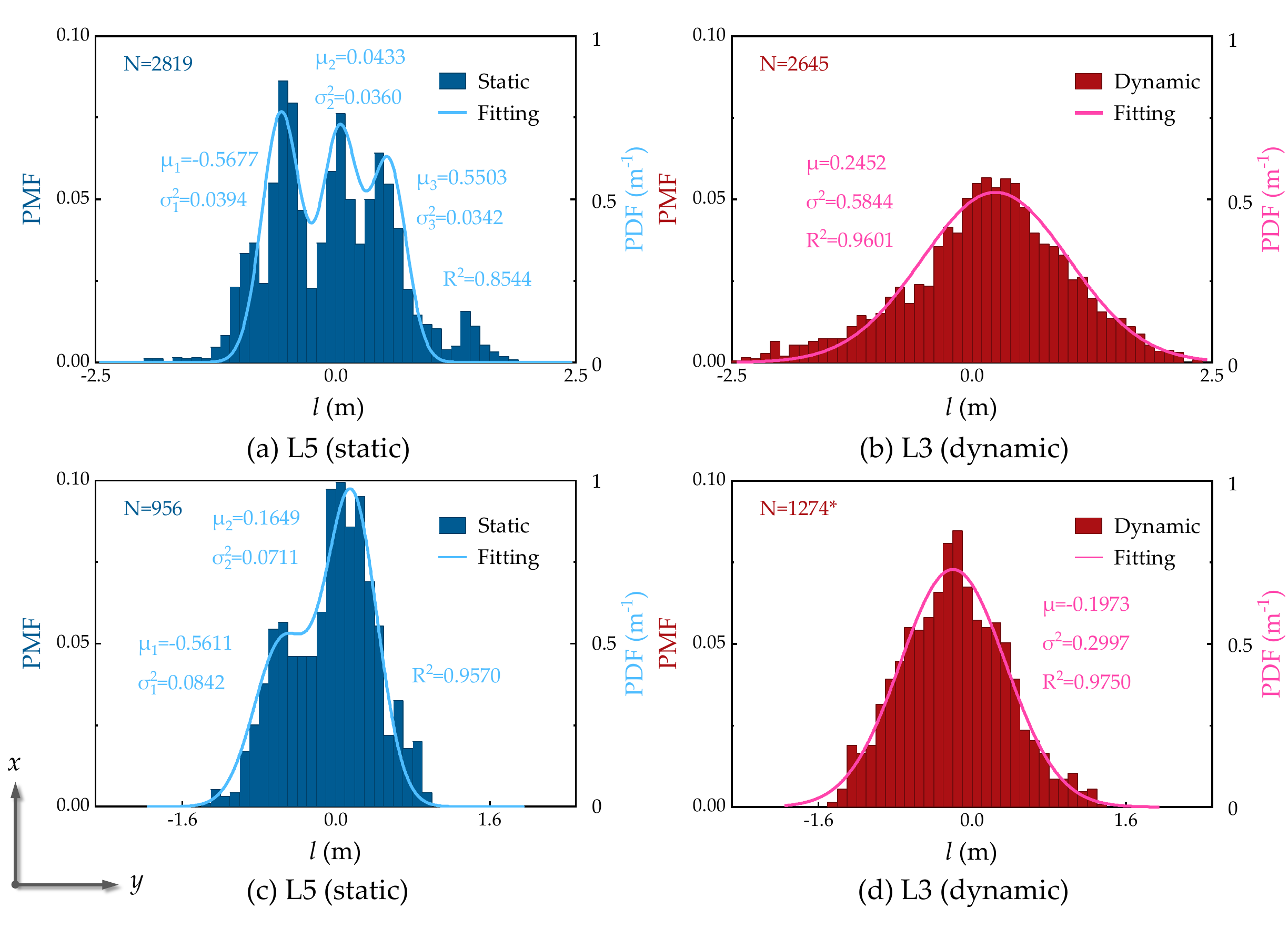}
\caption{Probability Mass Function (PMF) (bin size = 0.1 m) and Probability Density Function (PDF) of the location distribution (\(l\)), along with the corresponding Univariate Gaussian curves (Dynamic condition) and Gaussian mixture curve (Static condition), and the respective expectation (\(\mu\)) and variance (\(\sigma^2\)), are presented in the subplots and Appx. \ref{Appendix A}. In the figures, "\(N\)" denotes the sample size; * indicates that, according to the Anderson-Darling test, the data significantly originate from a normal distribution at the 0.05 level.}
\label{fig1}
\end{figure}

\begin{figure}[ht!]
\centering
\includegraphics[scale=0.35]{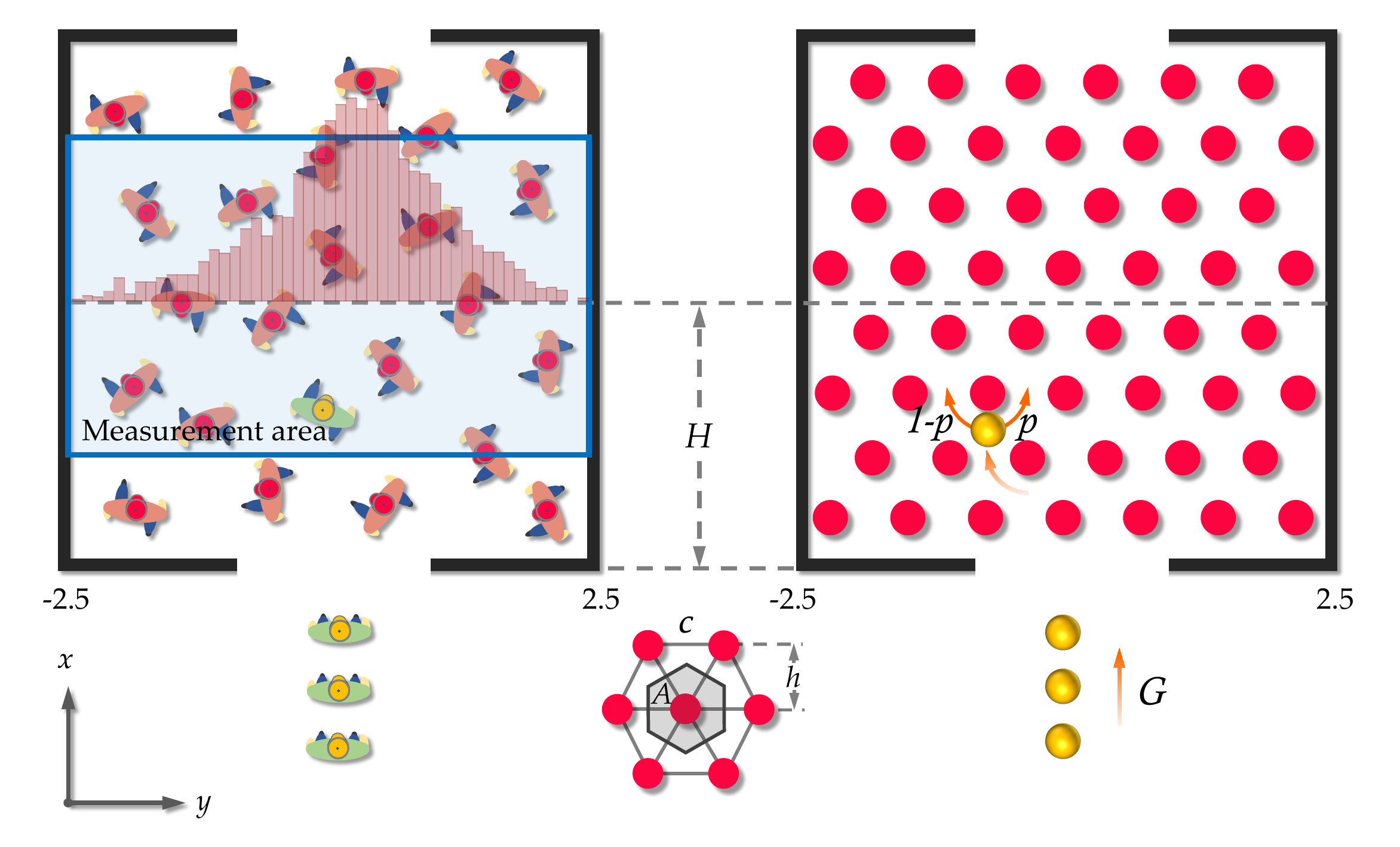}
\caption{Schematic illustration of the Galton board during the crossing process, wherein the pedestrians' motion is approximated akin to the descent of balls.}
\label{fig2}
\end{figure}

In the pedestrian crossing experiment (see \citet{wang2023}), we conducted a systematic assessment of the location distribution when pedestrians crossing the measurement area. The \(y\)-axis projection of pedestrian locations was subjected to statistical analysis and Fig. \ref{fig1} visualized the distribution of pedestrian locations within the measurement area. The distribution illustrated distinct characteristics within the crossing process. Statistical results indicate that pedestrian positions are concentrated at specific locales within the static context, characterized by the multi-peak distribution resulting from the phenomenon of cross-channel formation \citep{VideoExample}. In contrast, in dynamic contexts, due to the influence of stochastic effects, the spatial distribution of pedestrians tends to present a bell curve, as depicted in Fig. \ref{fig1}. We conjecture that this crossing process may be approximated to the descent of balls within a gravitational field, similar to the dynamics of the Galton board, as illustrated in Fig. \ref{fig2}. In line with this idea, we propose using the Galton board approximation to estimate the walking preferences of pedestrians within the crossing process.

\section{Method} \label{section2}
Based on the statistical results presented above, we can obtain the distribution function of pedestrian positions during the crossing process. Within dynamic conditions, the fit based on Univariate Gaussian curves allows us to obtain the mean and variance of pedestrian position distributions (the mean and variance present the complete information of the Univariate Gaussian distribution). Using this data, combined with the Galton board approximation, we can estimate the walking preferences of pedestrians.
We assume that participants are distributed uniformly within the experimental area. By harnessing global density metrics, an approximation of peg layers is derived, as illustrated in Fig. \ref{fig2} and computed according to Eq. \ref{1}. The average densities for experiments under low and high-density conditions are 1 ped/m\(^2\) and 2.47 ped/m\(^2\), respectively (see \citet{wang2023}). Subsequently, the corresponding peg layers can be ascertained through these designated average densities.

\begin{equation}\label{1}
\left\{ \begin{array}{l}
A = \frac{{2\sqrt 3 {h^2}}}{3} = \frac{{\sqrt 3 {c^2}}}{2}\\
n = \frac{H}{h}
\end{array} \right.
\end{equation}

Here, \(A\) represents the polygon area occupied by the individual, \(c\) denotes the distance between adjacent pegs, \(H\) signifies the crossing distance of pedestrians, \(h\) corresponds to the layer height of the Galton board, and \(n\) indicates the layers of the approximated Galton board.

According to Eq. \ref{1}, we can calculate the layers of the corresponding Galton boards within the low-density and high-density experiments, which are 2.6864 and 3.2069, respectively. Within the \(n\)-layer Galton board, the distribution of ball positions can be calculated via the corresponding equivalent binomial distribution \(B (n, p)\). Here, \(p\) denotes the probability of the ball moving right after a collision, and representing the right-side preference of pedestrians. Based on the statistical data presented in Fig. \ref{fig1} and the basic properties of the binomial distribution (see Appx. \ref{Appendix B}), we can derive the corresponding dimensionless equation:

\begin{equation}\label{2}
\left\{ \begin{array}{l}
\frac{\mu }{c} = np - \frac{n}{2}\\
\frac{{{\sigma ^2}}}{{{c^2}}} = np(1 - p)
\end{array} \right.
\end{equation}

The left term of Eq. \ref{2} represents the nondimensionalized mean and variance. Owing to the systematic rightward shift of the horizontal axis, as illustrated in Fig. \ref{fig1} (where the shift amount after nondimensionalization is \(n\)/2), the \(n\)/2 term on the right of the equation includes the corresponding compensation.

\subsection{Constant \(n\)} \label{subsection2.1}

Under ideal conditions, within an \(n\)-level Galton board, the ball will experience \(n\) collisions as it descends, which corresponds to pedestrians detouring \(n\) times in the crossing process. According to Eq. \ref{2}, under a constant \(n\) condition (\(n\)=2.6864 in the low-density experiment; \(n\)=3.2069 in the high-density experiment), the functions are unsolvable, hence it is impossible to obtain precise results regarding \(n\). A potential method is to estimate the pedestrian's right preference based on both the mean and the variance. However, since the mean and variance cover only partial information about the original distribution, estimating \(p\) solely based on individual parameters can lead to the following issues:

(1) Inconsistent evaluation results;

(2) Unpredictable errors in evaluation results (in some cases, the presence of imaginary roots, see Appx. \ref{Appendix c}).

In this consideration, use the estimated distribution to approximate \(p\) instead of relying on the original distribution maybe a feasible method. Following this idea, an objective function \(I\), can be established to represent the proximity between the estimated and original distributions. \(I\) is defined as the ratio of the intersection to the union of the original and estimated distributions. Hence \(I\in [0,1]\).
The objective function (Likelihood function) is:

\begin{equation}\label{3}
\max I=\frac{\int_{-\infty }^{\infty }{\min \left[ {{f}_{1}}(g(\mu ),h({{\sigma }^{2}})),{{f}_{2}}(\hat{g}(\mu ),\hat{h}({{\sigma }^{2}})) \right]dx}}{\int_{-\infty }^{\infty }{\max \left[ {{f}_{1}}(g(\mu ),h({{\sigma }^{2}})),{{f}_{2}}(\hat{g}(\mu ),\hat{h}({{\sigma }^{2}})) \right]dx}}.
\end{equation}

In Eq. \ref{3}, \({{f}_{1}}(g(\mu ),h({{\sigma }^{2}}))\) and \({{f}_{2}}(\hat{g}(\mu ),\hat{h}({{\sigma }^{2}}))\) represent the original distribution function and the estimated distribution function, respectively. The detailed expressions are as follows:

\begin{equation}\label{4}
\left\{ \begin{array}{l}
{f_1}(g(\mu ),h({\sigma ^2})) = \frac{1}{{\sqrt {2\pi {h^2}\left( {{\sigma ^2}} \right)} }}{e^{ - \frac{{{{\left( {x - g(\mu )} \right)}^2}}}{{2{h^2}\left( {{\sigma ^2}} \right)}}}},g\left( \mu  \right) = \frac{\mu }{c},h\left( {{\sigma ^2}} \right) = \frac{{{\sigma ^2}}}{{{c^2}}}\\
{f_2}(\hat g(\mu ),\hat h({\sigma ^2})) = \frac{1}{{\sqrt {2\pi {{\hat h}^2}\left( {{\sigma ^2}} \right)} }}{e^{ - \frac{{{{\left( {x - \hat g\left( \mu  \right)} \right)}^2}}}{{2{{\hat h}^2}\left( {{\sigma ^2}} \right)}}}}
\end{array} \right..
\end{equation}

Here, \(g\left( \mu  \right)\) and \(h\left( {{\sigma }^{2}} \right)\) denote the non-dimensional mean and variance of the original normal distribution, while \(\hat{g}\left( \mu  \right)\) and \(\hat{h}\left( {{\sigma }^{2}} \right)\) represent the maximum likelihood estimator (non-dimensional mean and variance) of the estimated distribution, respectively. The corresponding constraints are:

\begin{equation}\label{5}
s.t.\left\{ \begin{array}{l}
\hat g(\mu ) = np - \frac{n}{2}\\
\hat h({\sigma ^2}) = np(1 - p)\\
p \in [0,1]
\end{array} \right..
\end{equation}

\begin{figure}[ht!]
\centering
\includegraphics[scale=0.6]{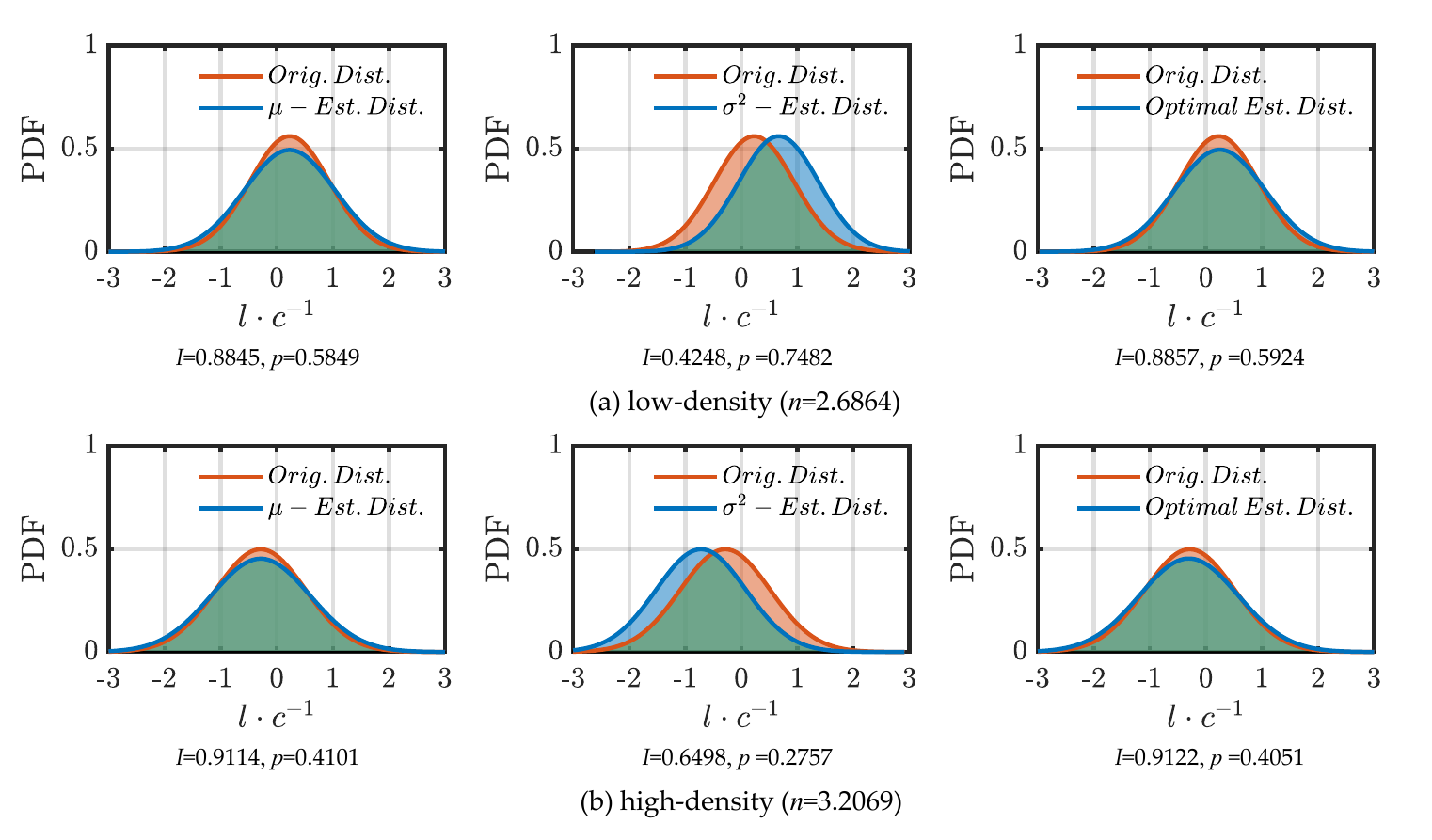}
\caption{Original and estimated distributions of pedestrian location within the high-density and low-density experiments.}
\label{fig3}
\end{figure}

By solving the objective function, we can obtain the optimal distribution closest to the original condition under the constraint-satisfying condition, along with the corresponding \(p\). The estimation results are illustrated in Fig. \ref{fig3}, where the likelihood method of \(\mu\) estimate, \(\sigma^2\) estimate, and the optimal estimate are compared.

Based on the optimal estimation, we obtained the corresponding estimates for the right-side preference probability in low-density and high-density experiments, which are \(p\)=0.5924 and \(p\)=0.4051, respectively.

\subsection{Variable \(n\)} \label{subsection2.2}

In practical situations involving ball descent and pedestrian motion, the frequency of collisions or detours is inherently uncertain due to the influence of stochastic effects. Therefore, for the Galton board model discussed in this paper, it is necessary to establish a variable \(n\) to estimate \(p\). In line with this, Eq. \ref{6} provides the corresponding quadratic equation in terms of the variables \(n\) and \(p\) based on Eq. \ref{2}, and satisfying \(n\ge 0,p\in [0,1]\).

\begin{equation}\label{6}
\left\{ \begin{array}{l}
\frac{{{n^2}}}{4} - \left( {\frac{\mu }{c} + \frac{{{\sigma ^2}}}{{{c^2}}}} \right)n + \frac{{{\mu ^2}}}{{{c^2}}} = 0\\
 - \frac{\mu }{c} \cdot {p^2} + \left( {\frac{\mu }{c} - \frac{{{\sigma ^2}}}{{{c^2}}}} \right) \cdot p + \frac{{{\sigma ^2}}}{{2{c^2}}} = 0
\end{array} \right.
\end{equation}

\begin{figure}[ht!]
\centering
\includegraphics[scale=0.3]{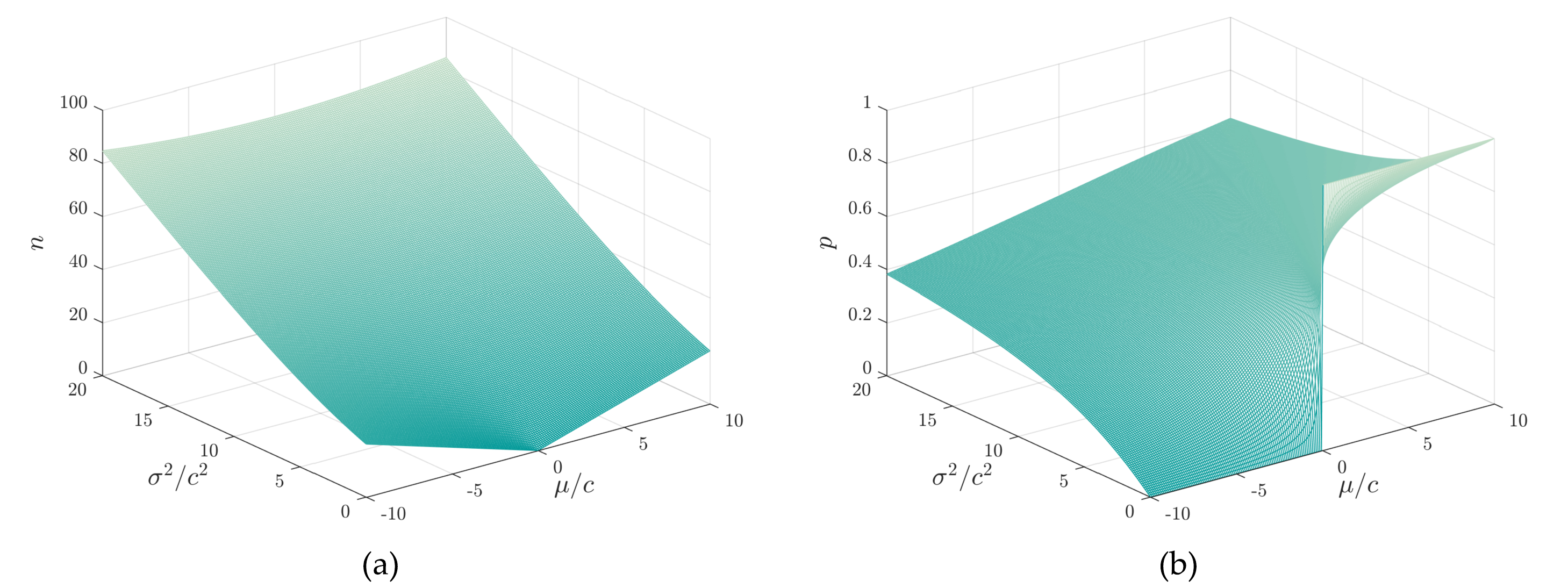}
\caption{Solutions for the variables \(n\) and \(p\) under various initial parameters.}
\label{fig4}
\end{figure}

From Eq. \ref{6}, it can be deduced that for any real inputs \(\mu\) and \(\sigma\), real roots for \(n\) and \(p\) always exist within the feasible domain. Fig. \ref{fig4} demonstrates the results for variables \(n\) and \(p\) under various initial distributions.

Within the given parameters, we can calculate the results of variables \(n\) and \(p\) in the corresponding low-density and high-density experiments (under the constraints of Eq. \ref{2}, the unbiased estimators for \(n\) and \(p\)):

\begin{equation}\label{7}
\left\{ \begin{array}{l}
n = 2.1225,p = 0.6075\\
n = 2.6852,p = 0.3926
\end{array} \right..
\end{equation}

The computational results are very close to those obtained under the assumption of a constant \(n\). The evaluated collision counts \(n\) in both high-density and low-density experiments are lower than the theoretical predictions, where \(n\) is assumed constant. These results suggest that pedestrians may engage in fewer detours during the crossing process compared to the theoretically conjectured detouring frequency.

\section{Conclusion} \label{section3}

We propose a method that utilizes the Galton board approximation to estimate pedestrians' walking preferences. The right-side preference probability was estimated under two conditions: constant \(n\) and variable \(n\). In the constant \(n\) condition, we solved by approximating the original distribution with an estimated distribution. In the variable \(n\) scenario, the analytical results indicate that solutions always exist for any given real parameters \(n\) and \(p\). Utilizing data from low-density and high-density experiments, and considering both the constant and variable \(n\) conditions, we calculated the probability of pedestrians' right-side preference, with results closely aligning. This method provides a reference for estimating pedestrian walking preferences in dense crowds.

It should be noted that this research aims to explore the possibility of utilizing the approximation of the Galton board to assess pedestrians' walking preferences within dynamic crowds. Due to deficiencies in the experimental  setup and potential measurement errors, the results obtained should not be considered as any type of empirical support for pedestrians’ walking preferences.

\section*{Data Availability}
The data can be found here: {\url{https://doi.org/10.34735/ped.2019.4}} (Pedestrian Dynamics Data Archive)
 or 
{\url{https://drive.google.com/drive/folders/1NYVnRp0z8VPuskfezMr51gB-sraOf6Iq?usp=drive_link}} (Google Drive).

\bibliographystyle{aasjournal}

\appendix
\section{Function of Gaussian Mixture Curve.} \label{Appendix A}
Tab. \ref{table1} and Tab. \ref{table2} presented the fitting parameters of the gaussian mixture curve corresponding to Fig. \ref{fig1} (a) and Fig. \ref{fig1} (c), respectively.

\begin{table}[ht]
\centering
\caption{Coefficients result from fitting the Gaussian Mixture curve to the results of the experiment (L5 class, static condition).}
\begin{tabular}{@{}cccccccccc@{}}
\toprule
\textbf{Function} & \multicolumn{9}{c}{\textbf{Parameters}} \\ 
\midrule
\multirow{2}{*}{$f(x)=\sum_{i=1}^{3}{\pi_i \cdot \frac{1}{\sqrt{2\pi \sigma_i^2}} e^{-\frac{(x-\mu_i)^2}{2\sigma_i^2}}}, \pi_i \geq 0, \sum_{i=1}^{3}{\pi_i=1}$} & $\pi_1$ & $\mu_1$ & $\sigma_1^2$ & $\pi_2$ & $\mu_2$ & $\sigma_2^2$ & $\pi_3$ & $\mu_3$ & $\sigma_3^2$ \\ 
 & 0.381 & -0.5677 & 0.0394 & 0.337 & 0.0433 & 0.0360 & 0.282 & 0.5503 & 0.0342 \\ 
\bottomrule
\end{tabular}
\label{table1}
\end{table}

\begin{table}[ht]
\centering
\caption{ Coefficients result from fitting the Gaussian Mixture curve to the results of the experiment (L3 class, static condition).}
\begin{tabular}{@{}ccccccc@{}}
\toprule
\textbf{Function} & \multicolumn{6}{c}{\textbf{Parameters}} \\ 
\midrule
\multirow{2}{*}{$f(x)=\sum_{i=1}^{2}{\pi_i \cdot \frac{1}{\sqrt{2\pi \sigma_i^2}} e^{-\frac{(x-\mu_i)^2}{2\sigma_i^2}}}, \pi_i \geq 0, \sum_{i=1}^{2}{\pi_i=1}$} & $\pi_1$ & $\mu_1$ & $\sigma_1^2$ & $\pi_2$ & $\mu_2$ & $\sigma_2^2$ \\ 
 & 0.364 & -0.5611 & 0.0842 & 0.636 & 0.1649 & 0.0711 \\ 
\bottomrule
\end{tabular}
\label{table2}
\end{table}

\section{BINOMIAL DISTRIBUTION.} \label{Appendix B}

The probability mass function of the binomial distribution provides the probability of the random variable \(X\) achieving \(k\) successes in \(n\) Bernoulli trials:

\begin{equation}\label{8}
P(X = k) = \left( \begin{array}{l}
n\\
k
\end{array} \right){p^k}{(1 - p)^{n - k}}
\end{equation}

Where \(p\) represents the probability of success, from which the expected mean can be inferred:

\begin{equation}\label{9}
E(X) = np
\end{equation}

Under ideal conditions, the variance satisfies:

\begin{equation}\label{10}
Var(X) = np(1 - p)
\end{equation}

For a Galton board with a given number of layers \(n\), the theoretical distribution of the falling balls' positions has been satisfying the binomial distribution \(B (n, p)\).

\section{Estimating \(p\) based on the variance, considering the possibility of complex roots.} \label{Appendix c}

According to the properties of the variance of the binomial distribution \(B (n, p)\), it can be established that the variance is related to the success probability \(p\) of the Bernoulli trials.

\begin{equation}\label{10}
Var(X) \propto p(1 - p)
\end{equation}

\begin{figure}[ht!]
\centering
\includegraphics[scale=1]{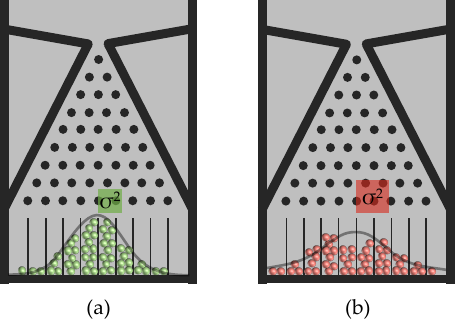}
\caption{Illustrations of ball distributions in the Galton board, where (a) presents the case with the maximum variance observed under ideal conditions, and (b) shows a scenario where the observed variance exceeds the ideal maximum variance.}
\label{fig5}
\end{figure}

Eq. \ref{10} illustrates the distribution law of the variance of the binomial distribution under ideal conditions, and the variance is maximized when \(p\) equals 1/2, as shown in Fig. \ref{fig5}(a). However, based on observation, stochastic effects may cause the data variance to exceed the theoretical maximum variance of the binomial distribution, as depicted in Fig. \ref{fig5}(b). 

\begin{figure}[ht!]
\centering
\includegraphics[scale=0.7]{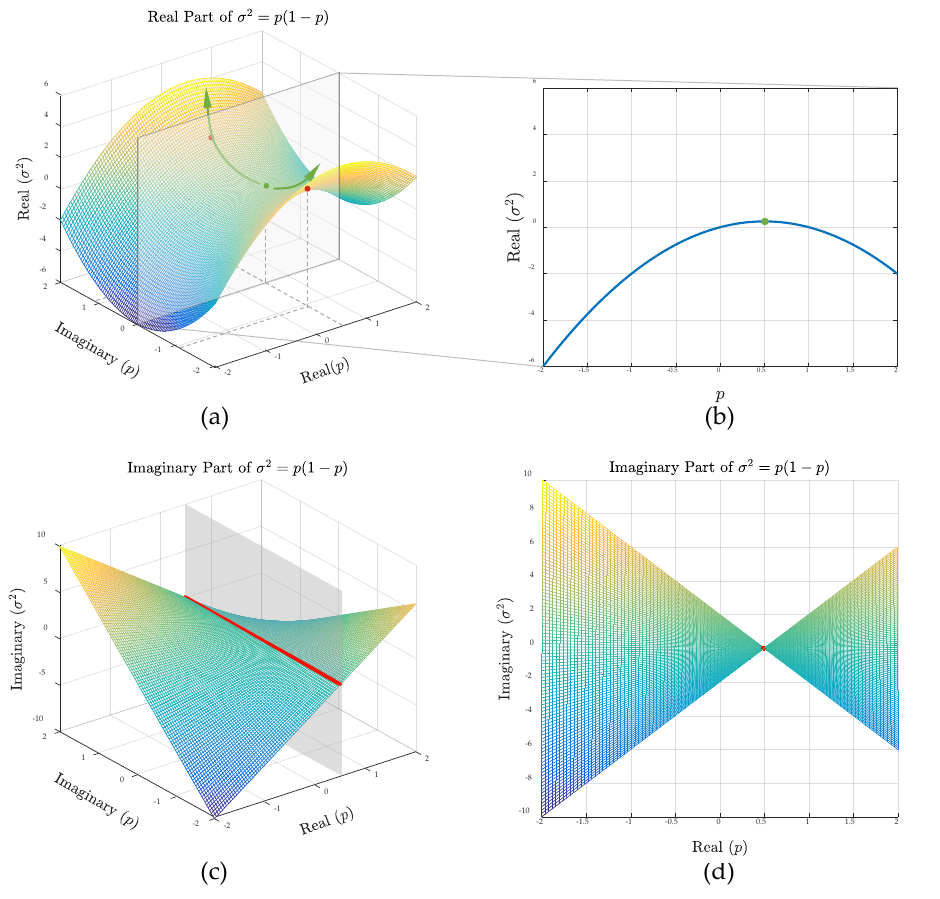}
\caption{Illustration of \({{\sigma }^{2}}=p(1-p)\) in the complex plan. subfigures (a) and (c) respectively illustrate the real and imaginary components of \({{\sigma }^{2}}\), while subfigures (b) and (c) present the corresponding sectional view and lateral perspective.}
\label{fig6}
\end{figure}

When the statistical variance of the data exceeds the theoretical maximum value of variance (the saddle point in Fig. \ref{fig6}(a), an assessment of \(p\) necessitates the introduction of the complex domain to obtain a theoretically larger variance, although such a probability does not exist in practice. The values of variance have a zero imaginary part on the plane, as illustrated in Fig. \ref{fig6}(c) and \ref{fig6}(d), aligning with the real nature of statistical data. Therefore, the extension of the imaginary part of \(p\) will be conducted on this plane. The assessment values of \(p\) corresponding to Fig. \ref{fig5}(a) and Fig. \ref{fig5}(b) are represented by the green point and the red points, respectively, in Fig. \ref{fig6}(a).

\end{document}